# Terahertz Twistoptics – engineering canalized phonon polaritons

*Maximilian Obst, Tobias Nörenberg, Gonzalo Álvarez-Pérez, Thales V. A. G. de Oliveira, Javier Taboada-Gutiérrez, Flávio H. Feres, Felix G. Kaps, Osama Hatem, Andrei Luferau, Alexey Y. Nikitin, J. Michael Klopf, Pablo Alonso-González, Susanne C. Kehr and Lukas M. Eng*

**The terahertz (THz) frequency range is key to study collective excitations in many crystals and organic molecules. However, due to the large wavelength of THz radiation, the local probing of these excitations in smaller crystalline structures or few-molecular arrangements, requires sophisticated methods to confine THz light down to the nanometer length scale, as well as to manipulate such a confined radiation. For this purpose, in recent years, taking advantage of hyperbolic phonon polaritons (HPhP) in highly anisotropic van der Waals (vdW) materials has emerged as a promising approach, offering a multitude of manipulation options such as control over the wavefront shape and propagation direction. Here, we demonstrate the first THz application of twist-angle-induced HPhP manipulation, designing the propagation of confined THz radiation between 8.39 and 8.98 THz in the vdW material alpha-molybdenum trioxide (α-MoO$_3$), hence extending twistoptics to this intriguing frequency range. Our images, recorded by near-field optical microscopy, show the frequency- and twist-angle-dependent change between hyperbolic and elliptic polariton propagation, revealing a polaritonic transition at THz frequencies. As a result, we are able to allocate canalization (highly collimated propagation) of confined THz radiation by carefully adjusting these two parameters, *i.e.* frequency and twist angle. Specifically, we report polariton canalization in α-MoO$_3$ at 8.67 THz for a twist angle of 50°. Our results demonstrate an unprecedented control and the manipulation of highly-confined collective excitations at THz frequencies, offering novel possibilities for nanophotonic applications.**

The terahertz (THz) frequency regime ($v = 0.1 – 10$ THz, the so-called THz gap), is key to a variety of recent novel applications, including the newest generation of mobile networks,[1,2] non-destructive sensing of organic molecules relevant for medical applications,[3,4] or even deep-space exploration.[5] With the growing technical progress on THz detectors and sources,[6,7] the THz regime is becoming more accessible for applications in science and technology, such that the investigation of material properties at THz frequencies and with nanometer resolution[8,9] has become an important branch in optics[10–12] and materials sciences.[13,14]

In nanotechnology the major challenge inherent to THz radiation arises from the large wavelengths ($\lambda \approx 30 – 3000\,\mu m$), which implies enormously-large footprints in THz optics (*i.e.* THz antennas[15]), and a typically weak light-matter interaction in small volumes ($< \lambda^3$). To circumvent these disadvantages and achieve energy and information transport at the nanoscale, strong confinement of THz radiation is fundamental, enabling *e.g.* single molecule research[16] or controlled wave propagation.[17] In that regard, polaritons elegantly provide high field confinement in sub-wavelength-sized devices[17] through the strong coupling between electromagnetic radiation and collective excitations in matter. Specifically at mid-infrared (MIR) to THz wavelengths, hyperbolic phonon polaritons (HPhP) in van der Waals (vdW) materials, *e.g.* in hexagonal boron nitride (hBN),[18] α-molybdenum trioxide (α-$MoO_3$),[19,20] calcium carbonate ($CaCO_3$),[21] β-gallium oxide (β-GaO)[22] and α-germanium monosulfide (α-GeS),[23] have been demonstrated to offer an extremely high spatial confinement, while additionally featuring low losses and long lifetimes.[18,19] Moreover, HPhPs with in-plane hyperbolic propagation, as hosted for instance by α-$MoO_3$ and α-GeS, naturally offer a highly directional energy and information transport, a critical element in the design of nanoscale optical devices.[19,20] As the propagation properties of HPhPs are directly dependent on the permittivity of both the vdW material itself and its environment, exploring tuning mechanisms might be highly advantageous for prospective applications.

Therefore, the exploration of ways to tune the propagation of HPhPs has become a major focus of scientific research in recent years, *e.g.* by the formation of meta-surfaces,[24,25] coupling with graphene plasmons,[26–28] or the use of different dielectric environments.[29] A particularly promising route, as reported recently, is stacking vdW layers at specific twist angles $\theta$ to form so-called "twisted bilayers" (TBLs) or even triple-layered structures, which allows adjusting the HPhP iso-frequency curves (IFCs) from hyperbolic to elliptic via manipulation of both twist angle and frequency.[30–34] In particular, at the transition point between the two types of IFCs, which corresponds to a change of topology,[35,36] the IFCs become flat and consequently the HPhP propagation is diffraction-less or "canalized",[30,33,37] thus providing a unique way to transport energy and information along one specific spatial direction. Nevertheless, the observation of this effect so far was limited to the narrow MIR spectral range corresponding to the hyperbolic Reststrahlenband (RB) of α-MoO$_3$, and its application to the THz spectral region was still missing.

In the present work, we demonstrate control of the propagation of HPhPs at THz frequencies by expending the twistoptics concept, *i.e.* the manipulation of the phononic responses of a vdW material (here α-MoO$_3$) by forming TBL structures, to the THz spectral range. In particular, we demonstrate canalization of THz HPhPs, accompanied by high THz field confinement, and observe the associated reduction of losses.[30] To study these phenomena, we apply scattering-type scanning near-field optical microscopy (s-SNOM) in combination with a tunable free-electron laser (FEL)[20,38] allowing for the direct visualization of the transition from elliptical to hyperbolic polaritonic IFCs in α-MoO$_3$ TBLs at THz frequencies with nanometer spatial resolution (see Fig. 1a), therefore succeeding in extending the application of twistoptics to the THz. We specifically illustrate the canalization of THz HPhPs, characterized by a highly collimated propagation and a very low intensity spread angle of ≤ 4°, at a frequency of $v_C$ = 8.67 THz and a twist angle of $\theta_C$ = 50°. Furthermore, we observe a confinement-factor $B_{exp.} = \frac{\lambda_0}{\lambda_{HPhP}} \cong 5$ and a considerably smaller near-field damping of THz canalized HPhPs as compared to THz HPhPs in layers of α-MoO$_3$.

**α-MoO$_3$ TBLs for THz polaritonics.** α-MoO$_3$ is a highly anisotropic, layered semiconducting material featuring an orthorhombic crystal structure (space group Pnma) constituted by different bonds along the main crystallographic directions: while the molybdenum and oxygen atoms are strongly covalent bonded within each layer, adjacent layers are bound by weak vdW forces along the z = [010] direction. Thus, the lattice constants a = 3.76 Å, b = 3.97 Å, and c = 14.43 Å are markedly different, attesting to the crystal's high anisotropy.[19] Consequently, the phononic response of the crystal lattice is strongly directional and anisotropic as well. In fact, the permittivity tensor $\hat{\varepsilon}$ of α-MoO$_3$ reveals distinct so-called "hyperbolic regimes", in which the permittivities along different directions have opposite signs.[20] Specifically in the THz regime, in-plane hyperbolic responses span from $v$ = 7.86 – 11.01 THz along the [001] direction (Reststrahlenband RB$_{001}$) and from $v$ = 11.01 – 11.70 THz along the [100] direction (Reststrahlenband RB$_{100}$; see Note S.1 in the Supporting Information).[20] In the following, we focus on the lower and spectrally broader RB$_{001}$, since we expect a stronger HPhP-visibility when using a light source of a non-zero bandwidth.[20] We note, however, that a similar HPhP dispersion engineering may technically be performed in the upper THz RB$_{100}$ as well.

For in-plane hyperbolic materials with in-plane permittivities Re($\varepsilon_{100}$) > 0 and Re($\varepsilon_{001}$) < 0, the related individual layer HPhP IFCs within angular sectors are defined by an opening angle $2\beta$ in *k*-space that is given by $\beta = \tan^{-1}\left(\sqrt{-\frac{\varepsilon_{001}}{\varepsilon_{100}}}\right)$.[30–33] Stacking two thin layers of α-MoO$_3$ results in the coupling of their individual HPhP modes, and thus, in the modification of the overall HPhP IFCs as a function of twist angle (see Fig. 1a). At low twist angles ($\theta <$ $|180° - 2\beta|$ for $\beta > 45°$; see Fig. 1b) the individual layer IFCs overlap and form two anti-crossing points (ACP) $N_{ACP}$, with the overall HPhP IFCs remaining hyperbolic.[30] At larger twist angles ($\theta > |180° - 2\beta|$ for $\beta > 45°$; see Fig. 1f) the individual layer IFCs overlap at four points, increasing $N_{ACP}$ to four.[30] Likewise, due to the frequency-dependent permittivity, the same topological change can be observed for a fixed twist angle $\theta$ when spectrally sweeping from the LO to the TO phonon frequency in the respective Reststrahlenband, hence increasing the absolute value of $\varepsilon_{001}$ and thereby $\beta$ (see Figs. 1c,e). Due to the increase of $N_{ACP}$ this change in the overall HPhP IFCs constitutes a topological transition.[30] At the critical angle $\theta_C = |180° - 2\beta|$, the overall IFCs become straight and parallel, resulting in a diffraction-less, canalized polariton propagation (see Fig. 1a,d).[30,33] The resulting regimes, featuring an effective hyperbolic and elliptic HPhP propagation separated by the critical angle $\theta_C(\upsilon)$, are displayed in Fig. 1a versus frequency $\upsilon$ within RB$_{001}$, with the color-scale indicating the expected THz confinement as extracted from transfer-matrix simulations. Particularly, in the frequency range $\upsilon = 8.39 – 8.98$ THz investigated in this work, theory predicts a canalized HPhP propagation at the critical angles $\theta_C = 38 – 62°$ with a confinement $B_{\text{theory}}$ between 4.4 and 9.

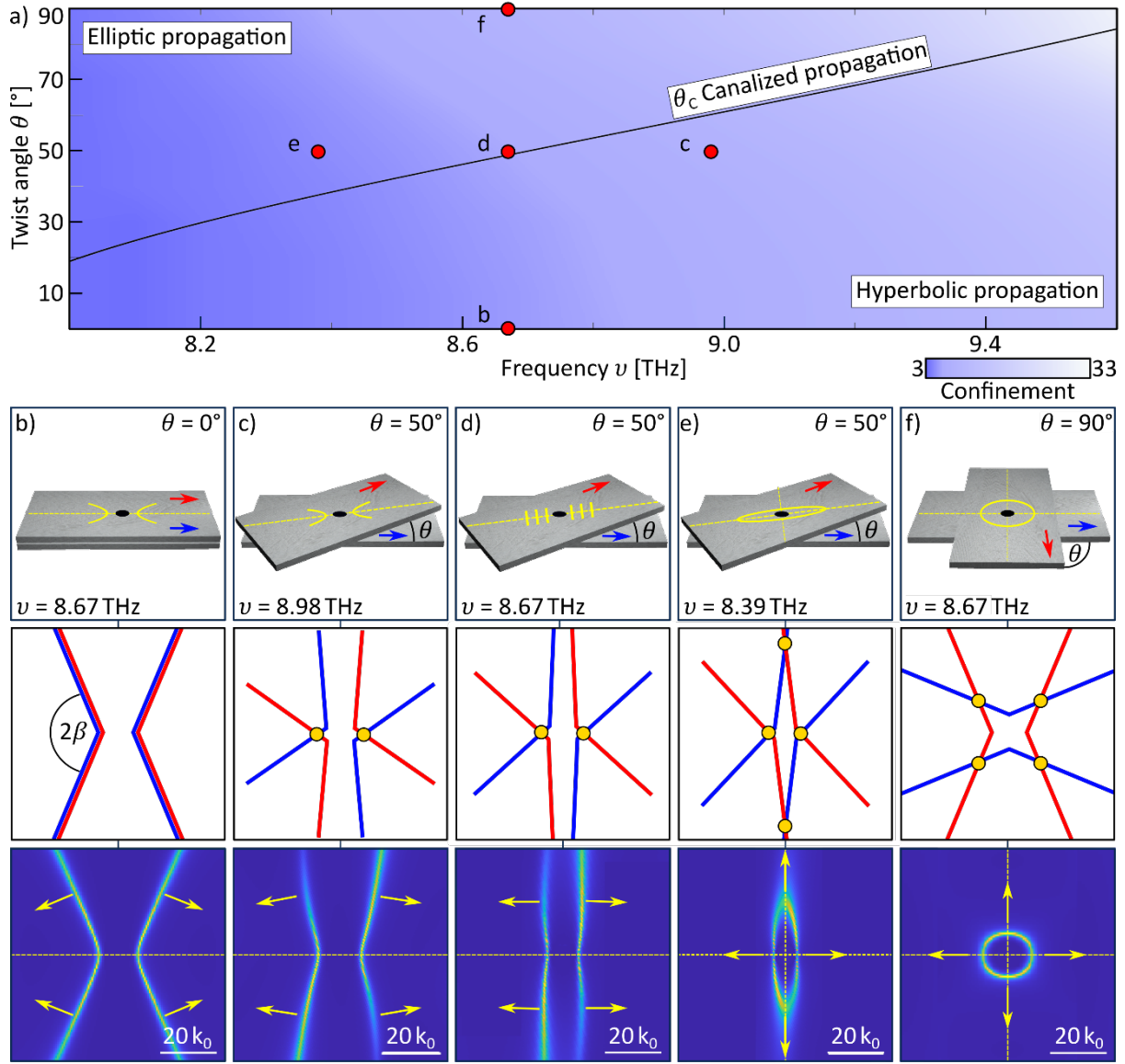

**Figure 1.** Twistoptics at THz frequencies: Topological HPhPs and canalization in TBLs. (a) Twist angle $\theta_C$ for canalized HPhPs propagation in the THz range as a function of frequency. The line separates elliptic and hyperbolic propagation regimes.[30] The color scale depicts the predicted polariton confinement, extracted from transfer-matrix calculations[39] assuming a stack of two 100nm thick layers of α-MoO$_3$. (b-f) Detailed overview of five representative twist angle-frequency pairs examined in this work, with (from top to bottom) schematic depictions, indications of individual layer IFCs and effective HPhP IFCs calculated by transfer-matrix formalism.[39] The black circles in the schematics represent HPhPs reflectors, the colored arrows indicate the [001] crystal directions of the top (red) and bottom (blue) layers. The individual IFCs follow the same color-scheme, with the yellow dots marking the anti-crossing points (ACP), whose number $N_{ACP}$ abruptly increases from 2 to 4 as the effective IFCs become elliptic. The yellow arrows in the simulations depict the direction of representative group velocities $\vec{v_g}$ defining the propagation direction (see yellow lines) and shape in real space.

## Results and Discussion

**THz polaritons with elliptic, hyperbolic, and canalized propagation.** To experimentally characterize the topological transition predicted at THz frequencies for HPhPs in α-MoO$_3$ TBLs (see Fig. 1a),[30] we perform polariton interferometry experiments (see methods) of three samples with twist angles of 0°, 50° and 90° (see methods), coupling s-SNOM with a FEL (see Fig. 2a) at seven selected frequencies within the THz RB$_{001}$, i.e. $v$ = 8.28, 8.39, 8.59, 8.67, 8.80, 8.98, and 9.38 THz. The recorded near-field S$_{2\Omega}$ images are evaluated both in real and k-space, and then compared to theoretical calculations obtained via transfer-matrix methods.[39] Selected key results are depicted in Figures 2 and 3 when varying both twist angle and frequency, respectively, while all other measurements as well as information about the propagation direction of the HPhPs are shown in Note S.2 in the Supporting Information.

Figure 2 shows the impact of the twist angle $\theta$ on the IFCs shape (as well as on the interference patterns seen in the real-space near-field images) for a constant frequency $v$ = 8.67 THz. At $\theta$ = 0° (Fig. 2b), the measured near-field image (left) exhibits a periodic pattern of signal maxima and minima indicating an in-plane hyperbolic propagation centered along the [001] direction, as expected for the case of a single layer (the bright circle at the center of the image results from measuring the bottom of the focused-ion-beam (FIB) milled hole that acts as a HPhP reflector, see methods). The 2D k-space map of the real-space measurement (middle panel, extracted by 2D fast-Fourier-transformation, FFT; see Note S.3 in the Supporting Information) exhibits a hyperbola, in excellent agreement with transfer matrix calculations (right panel),[39] thus corroborating the in-plane hyperbolic propagation of HPhPs. The observed k-vector "cutoff" in the experiment as compared to the calculations, stems from the experimentally reduced launching efficiency of large k-vectors resulting from the finite tip curvature,[40] as well as their reduced reflection at rounded edges of the hole.[41]

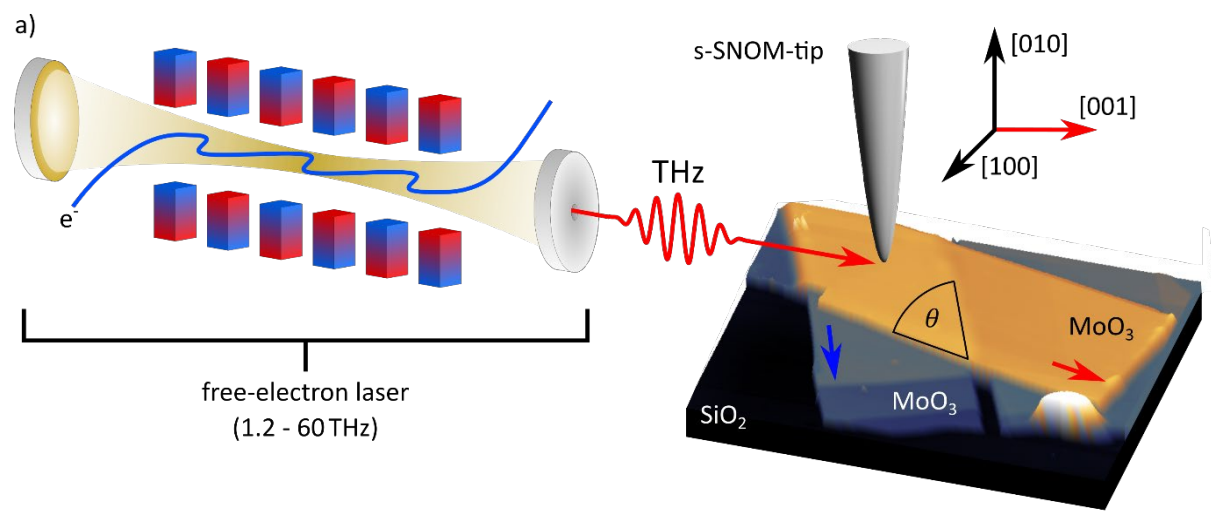

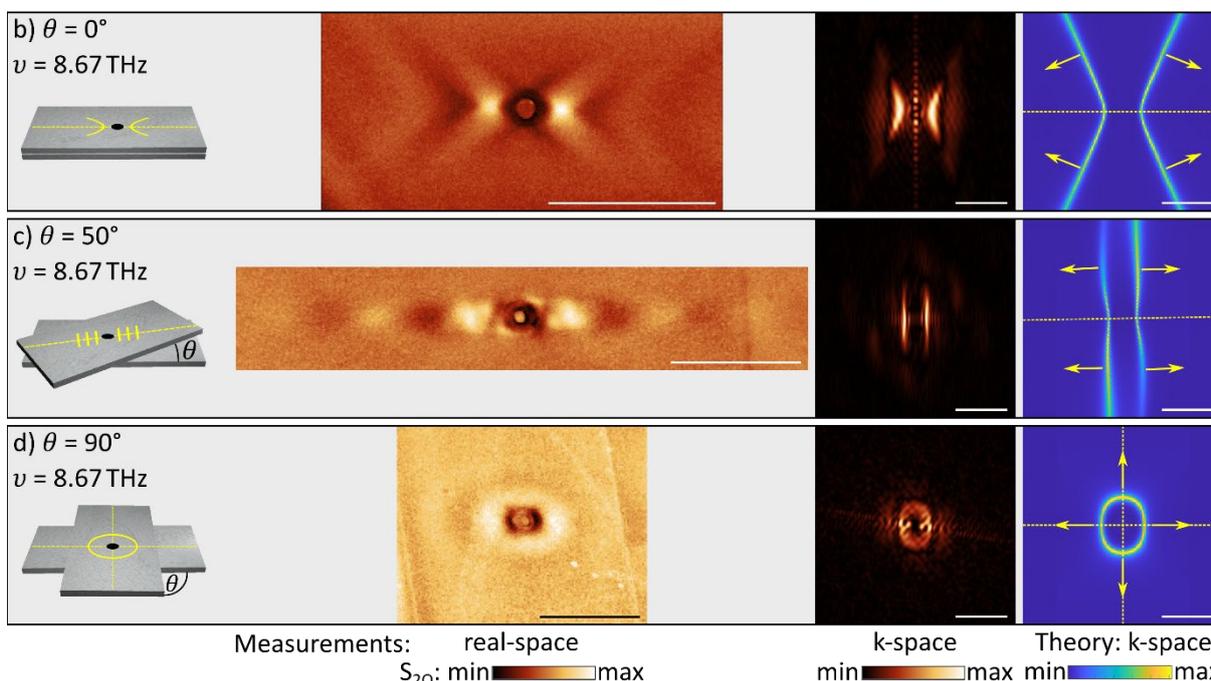

**Figure 2.** Visualization of topological THz HPhPs and canalization in α-MoO$_3$ TBLs as a function of twist angle $\theta$ ($v$ = const. = 8.67 THz). (a) Sketch of the free-electron-laser-based s-SNOM setup. The AFM image underneath the s-SNOM tip corresponds to the 3D topography of the TBL (twist angle $\theta$ = 50°). The coordinate system indicates the crystallographic axes of the α-MoO$_3$ flakes. The [001] crystallographic in-plane direction of the top and bottom flakes are marked by the red and blue arrows, respectively, corresponding to the cases given in Figs. 1b,d and f. (b-d) s-SNOM amplitude images taken at $v$ = 8.67 THz for $\theta$ = 0°, 50° and 90°, respectively. Sketches of the TBLs and experimental/theoretical IFCs are shown to the left and right, respectively. Scale bars: 5 μm for real-space, 40$k_0$ and 20$k_0$ for the experimental and theoretical IFCs, respectively.

At $\theta = 90°$ (Fig. 2d) the observed HPhP propagation is drastically different. As expected from theoretical considerations (see Figs. 1a,f), instead of the directed, hyperbolic propagation supported by the individual layers, the HPhP propagate in-plane elliptic at this twist angle[a]. In-between these two observed kinds of propagation, at $\theta = 50°$ (Fig. 2c), yet another kind of propagation is observed: the HPhP fringes are only visible along a straight line extending from the hole, without any noticeable diffraction; a canalized propagation. The corresponding IFCs revealed by 2D FFT are straight and parallel, matching the theoretical predictions.

These experimental observations, in very good agreement to the theoretical predictions, confirm that the coupling of the individual-layer HPhP modes and the corresponding topological transition from hyperbolic to elliptic propagation, can be achieved in TBLs in the THz spectral range by changing the orientation of the individual-layer IFCs in regard to each other (see Figs. 1b,d,f). They also reveal canalized HPhP propagation in TBLs in the THz spectral range, and indeed mark the first time such a propagation has been measured in this frequency range. The features of this unique type of propagation will be discussed in more detail below and in Figure 4.

---

[a] The elliptic, non-circular shape of the propagation, opposed to the prediction in Fig. 1f for this twist angle, is the result of the non-equal thickness of the two stacked α-MoO$_3$ flakes (see methods and Note S.5 in the Supporting Information).

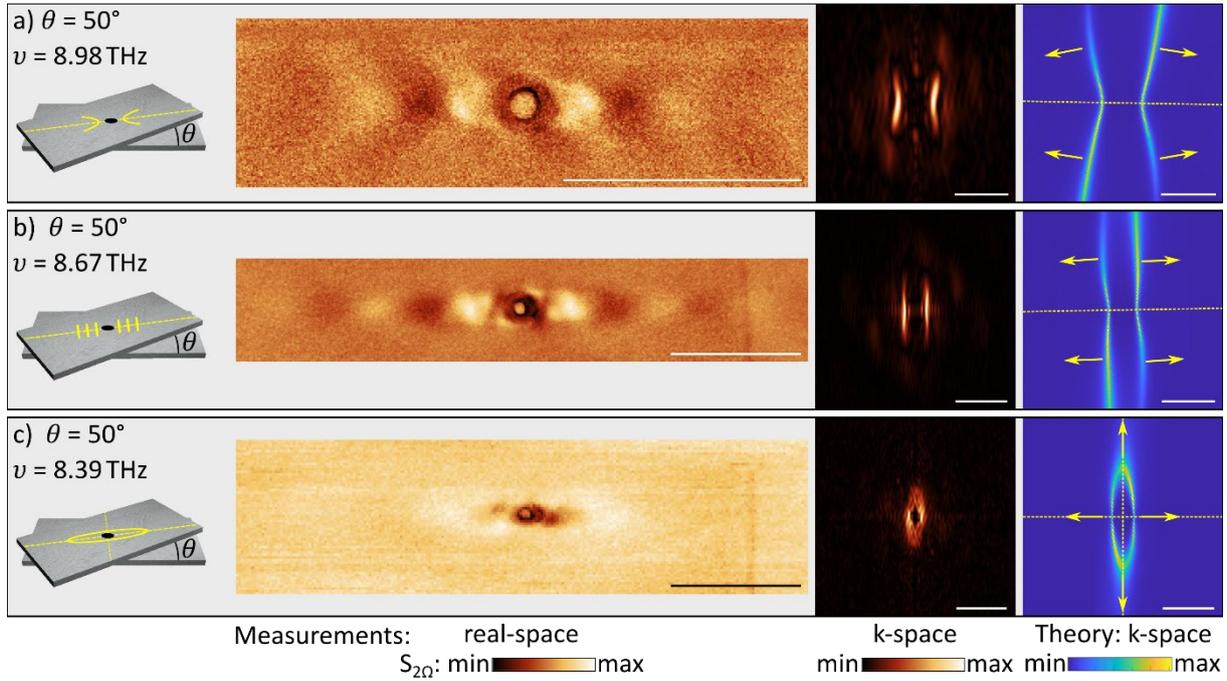

**Figure 3.** Visualization of topological THz HPhPs and canalization in α-MoO$_3$ TBLs as a function of incident frequency $v$ ($\theta$ = const. = 50°). (a-c) s-SNOM amplitude images taken at $\theta$ = 50° for $v$ = 8.98 THz, 8.67 THz and 8.39 THz, respectively, corresponding to the cases given in Figs. 1c,d and e. Sketches of the TBLs and experimental/theoretical IFCs are shown to the left and right, respectively. Scale bars: 5 µm for real-space, 40$k_0$ and 20$k_0$ for the experimental and theoretical IFCs, respectively.

However, while for different twist angles these changes in the propagation are induced by the change of the orientation of the individual flakes IFCs with regard to each other, for different frequencies this effect originates from the modified opening angles of the individual flake IFCs (see Fig. 1c,d,e), from $\beta$ = 90° near the TO mode to $\beta$ = 0° near the LO mode. These altered opening angles change the overall IFCs of the TBLs, giving rise to the effect of altered propagation as observed in our measurements.

In summary, the results in Figures 2 and 3 clearly display the control of the HPhP dispersion via both twist angle and frequency manipulation in the THz regime and show the possibility of achieving canalized HPhPs in the THz spectral range.

**THz HPhP characterization.** The above discussion features an excellent qualitative agreement of simulation and experiment. In order to further characterize the HPhPs, we now quantitatively compare the wavelength confinement $B = \lambda_0/\lambda_{\text{HPhP}}$ and quality factor $Q = \frac{\text{Re}(k)}{\text{Im}(k)}$ for the five measurements depicted in Figs. 2 and 3 (for more details on the data extraction and a full overview over the extracted values see Note S.4 in the Supporting Information), and compare these experimental results with data extracted from transfer-matrix simulations. For the elliptic propagation we extracted the $k$-values along the short axis of the ellipsis in $k$-space.

Among these five measurements we find the strongest confinement of $B_{\text{exp.}} = 8.3 \pm 1.6$ for a twist angle of $\theta = 0°$ and a frequency of $\nu = 8.67$ THz, compared to a theoretically expected confinement of $B_{\text{theory}} = 6.9$. Note that this comparably higher value for $B$ is related to the dependence of the HPhP wavelength on the sample thickness (see Note S.5 in the Supporting Information), which is for this particular sample thinner compared to the twisted stacks (see methods). When keeping the frequency constant but changing to a larger twist angle at $\theta = 50°$ (the sample for this angle has an increased overall thickness), the confinement factor is reduced to $B_{\text{exp.}} = 5.3 \pm 0.4$, in good agreement with the theoretically predicted $B_{\text{theory}} = 5.1$. Increasing the twist angle further to $\theta = 90°$ (while keeping the sample thickness) leads to approximately the same confinement of $B_{\text{exp.}} = 5.1 \pm 0.7$, which is significantly lower than the expected $B_{\text{theory}} = 10.6$. Overall, our experimental results imply that an increase of the twist angle reduces the confinement of the HPhPs, a tendency which is not mirrored in the theoretical findings based on the transfer-matrix calculations. One possible explanation for this phenomenon can be found in the experimental conditions: The rounded edges of the implemented HPhP scatterers reduce the reflection efficiency of HPhPs, an effect that increases with decreasing HPhP wavelengths[41]. This effect favors HPhPs with longer wavelengths and, hence, reduces the confinement.

A similar behavior is found when keeping the twist angle constant at $\theta = 50°$ while changing the frequency. In comparison to the values found at $v = 8.67$ THz, the confinement of the HPhP is increased for higher frequencies, at $v = 8.98$ THz to $B_{exp.} = 8.1 \pm 0.6$. This is confirmed by theory, although the predicted confinement $B_{theory} = 6.7$ is again lower than the measured one. In contrast, at a lower frequency of $v = 8.39$ THz the lowest confinement $B_{exp.} = 3.1 \pm 0.7$ has been found, in good agreement to theory, which predicts a confinement of $B_{theory} = 4.1$. This dependency of the HPhP wavelength and therefore the THz confinement on the frequency matches the one found in single layer measurements.[20]

While for confinement, theory and experiment are in relatively good agreement, the differences between the observed and predicted quality factor are drastic. In our measurements we find quality factors between 1.1 ($\theta = 0°$, $v = 8.67$ THz) and 2.9 ($\theta = 50°$, $v = 8.39$ THz), compared to a predicted range of 12.07 ($\theta = 90°$, $v = 8.67$ THz) to 16.7 ($\theta = 50°$, $v = 8.67$ THz). There are multiple explanations for the strongly reduced $Q$s. The utilized light source is likely the main source, as, firstly, a non-zero bandwidth strongly increases the observable decay length and, secondly, the signal-to-noise ratio that can be achieved using a FEL is not as high as for table-top light sources.[20,23,42] Additionally, the implantation of Ga-ions into the crystal structure and the subsequent sample-annealing (see Note S.6 in the Supporting Information) result in a reduction of the quality of the crystal structure, which corresponds to an increase of damping. In total these contributions result in a drastic underestimation of the $Q$ values. More accurate estimations of $Q$ would require a light source with a narrower bandwidth as well as the reduction of ion implantation, for example by using Helium- instead of Ga-ions for the FIB.

**Diffraction-less THz HPhP propagation at the topological angle.** Canalized HPhPs feature unique properties, such as exceptional directionality and a decreased decay over the propagated distance, which we quantify and discuss in the following. Experimentally, we observe such canalized THz HPhPs in a TBL α-MoO$_3$ system at a twist angle of $\theta = 50°$ and a frequency of $\upsilon = 8.67\,\text{THz}$ (see Figs. 2c, 3c, 4a), matching the predicted frequency of 8.70 THz.[30] Mathematically, the near-field signal containing the tip-launched and edge-reflected HPhP field amplitude Re($E_z$) at the flake surface, can be described by an exponentially decaying sine function[43]:

$$E_z(x) = E_{z_0} + \frac{A}{x^F} * \exp[-2\text{Im}(k)x]\sin[2\text{Re}(k)x - \varphi_c] \quad (1)$$

with $E_{z_0}$ corresponding to an offset due to the dielectric response of α-MoO$_3$, $A$ the amplitude of the HPhP at the position of the reflector, $x$ the distance from the HPhP reflector, $\varphi_c$ a phase offset, and $F$ the form factor accounting for the geometrical spreading of the HPhPs. In particular for hyperbolic and elliptical propagation, the form factor takes on the value $F = 0.5$,[30,33,43] whereas the diffraction-less propagation of a canalized HPhP entails $F = 0$. Note that the factor 2 both in the exponent and sine accounts for the double-path length when performing polariton interferometry, and thus equals 1 when fitting HPhPs launched by a separate launcher or sample structure instead of the tip.[20,23]

To demonstrate the significant effect of the form factor on the HPhP damping two transfer-matrix simulations were performed at $\upsilon = 8.67$ THz, one featuring canalized propagation matching the experimental conditions ($\theta = 50°$, $d_{\text{top}} = 125\,\text{nm}$, $d_{\text{bottom}} = 115\,\text{nm}$; Fig. 4b) and one featuring hyperbolic propagation ($\theta = 0°$, $d_{\text{total}} = 149\,\text{nm}$ adapted to feature similar HPhP wavelengths; Fig. 4c). In order to evaluate the damping, line profiles of the simulated HPhPs originating from a dipole-launcher were extracted along the propagation directions and fitted using formular (1) with a form factor $F = 0$ (see Fig. 4b) and $F = 0.5$ (see Fig. 4c), respectively. The envelopes of these fits $\frac{A}{x^F} * \exp[-\text{Im}(k)x]$ (see black and red dashed line on the right side of Fig, 4b,c) show that, due to the additional damping induced by the form factor, the amplitude of the HPhP with hyperbolic propagation decays much faster as compared to the amplitude of the HPhP with canalized propagation.

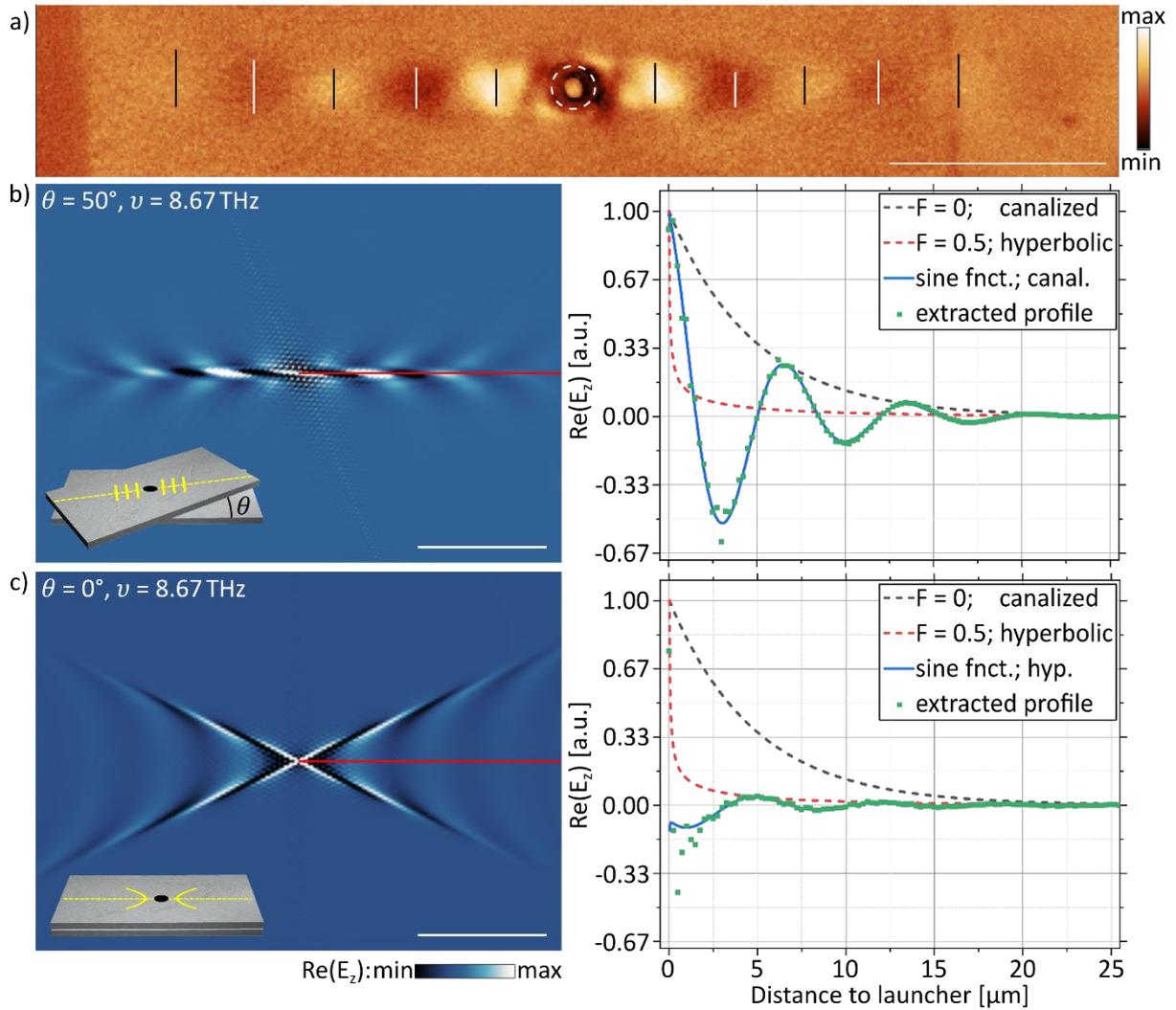

**Figure 4.** Canalized vs. hyperbolic THz HPhP propagation in α-MoO$_3$ TBLs. (a) Near-field image, S$_{2\Omega}$, taken at $\nu$ = 8.67 THz, showing canalized HPhPs in a twisted structure with $\theta_C$ = 50° and a frequency (see Fig. 2d). The black (white) vertical lines show the FWHM of the maxima (minima) extracted from a Gaussian fit. (b) Left: Analytical real-space simulation for the case shown in (a). Right: Fitted profiles (along the red line depicted in the simulated image) and envelopes for canalized and hyperbolic propagation. (c) Same as in (b) for a single flake with thickness $d_{total}$ = 149 nm that features hyperbolic HPhPs with similar wavelengths, thus allowing for a direct comparison. Scale bars: 5 μm in (a), 10 μm in (b) and (c)

Lastly, the directionality of the measured, canalized HPhP in Fig. 4a is evaluated by analyzing the full width at half maximum (FWHM) of the observed maxima and minima of its amplitude, via Gaussian fits perpendicular to the propagation direction. A strong lateral confinement of the THz HPhP with an experimentally obtained lateral HPhP spread which amounts to an opening angle of around α = 3.7° is found, thus confirming the exceptional directionality of energy transport.

**Conclusion**

In conclusion, we established the nanoscale manipulation of confined THz radiation by means of HPhPs in TBL stacks of α-MoO$_3$, thus extending the application of twistoptics to this frequency range. The resulting device shows overall HPhP IFCs that swap between hyperbolic and elliptic as a function of both twist angle between the two HPhP-hosting layers as well as frequency. We explored the degree of tunability at frequencies between 8.39 THz and 8.98 THz, and thereby demonstrated the canalized propagation of THz HPhPs. At the point of topological transition ($v$ = 8.67 THz, $\theta_C$ = 50°), we observe a confinement-factor of $B_{exp.}$ = 5.3 ± 0.4 as well as a lateral spread of the HPhPs that is reduced to an angle of 3.7°, which, as we have shown through theoretical examinations, results in a propagation with significantly reduced losses as compared to hyperbolic propagation. Our measurements mark the first time such a topological transition coupled with canalized propagation is observed in the THz spectral range, therefore extending the application of twistoptics to these lower frequencies. The resulting control via twistoptics over strongly confined THz radiation, which, as shown recently, can even be enhanced by stacking three instead of two layers,[34] opens the way to develop future nanophotonic devices operating at THz frequencies.

## Methods

**Sample Description.** In this work, we examine three different α-MoO$_3$-bilayer structures, each consisting of two about 100 nm-thick flakes stacked at the respective twist angles of $\theta = 0°$, 50°, and 90° (see Figs. 1b-d). Twisted stacks of α-MoO$_3$ slabs were fabricated with the dry transfer technique. We first exfoliated commercial α-MoO$_3$ bulk materials (Alfa Aesar) using Nitto tape (Nitto Denko Co., SPV 224P). Afterward, we performed a second exfoliation of the α-MoO$_3$ flakes from the Nitto tape to a transparent poly-(dimethylsiloxane) (PDMS) stamp, which was mounted in a micromanipulator for alignment and twisting. The second α-MoO$_3$ flake was precisely aligned, twisted at the desired angle, and released on top of the first flake, which was peeled off from the PDMS to a SiO$_2$ substrate in advance. Note that the mechanical exfoliation results in slight variations of the thicknesses ($d$) in different samples, namely $d_{total} = 125$ nm for $\theta = 0°$, $d_{top} = 125$ nm, and $d_{bottom} = 115$ nm for $\theta = 50°$, and $d_{top} = 78$ nm, and $d_{bottom} = 116$ nm for $\theta = 90°$. To provide a possibility for symmetric HPhP reflection in polariton interferometry, circular holes with a diameter of approximately 1 μm were milled/cut down to the substrate via FIB milling using Ga$^+$-ions. Note that the holes produced via this process have naturally rounded edges. Finally, to reduce crystal damage due to Ga-implantation, the samples were annealed at 350 °C for 3 hours. More details on the sample fabrication process are presented in Note S.6 of the Supporting Information.

**Polariton interferometry via THz-s-SNOM.** The visualization of highly confined THz HPhPs requires an experimental examination technique that provides sub-diffraction-limit optical resolution at THz wavelength, properties which have been shown to be achievable by the combination of the s-SNOM measurement technique with electron-accelerator based light sources.[20,44,45] Here, we combine s-SNOM with the free-electron laser FELBE at the Helmholtz-Zentrum Dresden-Rossendorf (HZDR), that offers narrowband THz radiation with an exceptional spectral tunability and high spectral power densities (see Fig. 1a and Note S.7 of the Supporting Information).[20,38,46] In s-SNOM, the high lateral optical resolution of about 20nm is defined directly by the nm-sized probe, which acts as an antenna to the incident illumination and provides an enhanced near-field focus at its apex,[47,48] which then enables the excitation of high-momentum phonon polaritons in the sample that propagate across the surface away from the tip. Such tip-launched HPhPs get reflected at the edges of the α-MoO$_3$ flake and in turn propagate back to the probe, where they are scattered into the far-field. Thus, by scanning the sample under the spatially fixed near-field probe, the obtained near-field images correspond to an interference pattern of the launched/back-reflected HPhPs and the directly back-scattered light at the position of the tip, enabling real-space visualization.[18,49,50] By reflecting the HPhP at artificially created holes in the flakes, which correspond to flake edges, the full in-plane propagation can be measured. Note that due to the HPhP travelling from tip to edge and back, the spacing of the maxima in the observed interference patterns relates to half the wavelength of the HPhPs.[43]

## Associated Content

### Supporting information

The supporting information is available free of charge at ...

Details on α-MoO$_3$ THz permittivity data, additional measurements at all mentioned frequencies, details on data processing, the extraction of HPhP confinements and quality factors from line profiles and an overview over extracted properties as well as a remark on the influence of sample thickness, sample preparation and FEL and setup parameters (PDF).

# Author Information

**Corresponding Authors**

**Maximilian Obst** – *Institut für Angewandte Physik, Technische Universität Dresden, Dresden 01187, Germany; Würzburg-Dresden Cluster of Excellence - EXC 2147 (ct.qmat), Dresden 01062, Germany;*
E-mail: maximilian.obst@tu-dresden.de

**Susanne C. Kehr** – *Institut für Angewandte Physik, Technische Universität Dresden, Dresden 01187, Germany; Würzburg-Dresden Cluster of Excellence - EXC 2147 (ct.qmat), Dresden 01062, Germany;*
E-mail: susanne.kehr@tu-dresden.de

**Authors**

**Tobias Nörenberg** – *Institut für Angewandte Physik, Technische Universität Dresden, Dresden 01187, Germany; Würzburg-Dresden Cluster of Excellence - EXC 2147 (ct.qmat), Dresden 01062, Germany;*

**Gonzalo Álvarez-Pérez** – *Department of Physics, University of Oviedo, Oviedo 33006, Spain; Center of Research on Nanomaterials and Nanotechnology CINN (CSIC‑Universidad de Oviedo), El Entrego 33940, Spain;*

**Thales V. A. G. de Oliveira** – *Institut für Angewandte Physik, Technische Universität Dresden, Dresden 01187, Germany; Würzburg-Dresden Cluster of Excellence - EXC 2147 (ct.qmat), Dresden 01062, Germany; Institute of Radiation Physics, Helmholtz-Zentrum Dresden-Rossendorf, Dresden 01328, Germany;*

**Javier Taboada-Gutiérrez** – *Section de Physique, Université de Genève, Genève 1211, Switzerland;*

**Flávio H. Feres** – *Gleb Wataghin Physics Institute, University of Campinas (Unicamp), Campinas, SP, Brazil; Brazilian Light Laboratory (LNLS), Brazilian Center for Research in Energy and Materials (CNPEM), Campinas, SP, Brazil;*

**Felix G. Kaps** – *Institut für Angewandte Physik, Technische Universität Dresden, Dresden 01187, Germany; Würzburg-Dresden Cluster of Excellence - EXC 2147 (ct.qmat), Dresden 01062, Germany;*

**Osama Hatem** – *Institut für Angewandte Physik, Technische Universität Dresden, Dresden 01187, Germany; Würzburg-Dresden Cluster of Excellence - EXC 2147 (ct.qmat), Dresden 01062, Germany; Department of Engineering Physics and Mathematics, Faculty of Engineering, Tanta University, Tanta 31511, Egypt;*

**Andrei Luferau** – *Institute of Ion Beam Physics and Materials Research, Helmholtz-Zentrum Dresden-Rossendorf, Dresden 01328, Germany; Institut für Angewandte Physik, Technische Universität Dresden, Dresden 01187, Germany;*

**Alexey Y. Nikitin** – *Donostia International Physics Center (DIPC), Donostia/San Sebastián 20018, Spain; IKERBASQUE, Basque Foundation for Science, Bilbao 48013, Spain;*

**J. Michael Klopf** – *Institute of Radiation Physics, Helmholtz-Zentrum Dresden-Rossendorf, Dresden 01328, Germany;*

**Pablo Alonso-González** – *Department of Physics, University of Oviedo, Oviedo 33006, Spain; Center of Research on Nanomaterials and Nanotechnology CINN (CSIC‑Universidad de Oviedo), El Entrego 33940, Spain;*

**Lukas M. Eng** – *Institut für Angewandte Physik, Technische Universität Dresden, Dresden 01187, Germany; Würzburg-Dresden Cluster of Excellence - EXC 2147 (ct.qmat), Dresden 01062, Germany; Collaborative Research Center 1415, Technische Universität Dresden, Dresden 01069, Germany;*

**Author contributions**

T.V.A.G.O. together with M.O., T.N., S.C.K., and L.M.E. initiated the research. J.T.-G., J.D. and P.A.-G. prepared the required samples. M.O and J.D. performed the pre-characterization of the samples. J.M.K., M.O., L.M.E., and S.C.K. prepared the instrumentation for the FEL measurements. M.O., T.N., F.H.F., F.G.K., O.H. and A.L. conducted the polariton interferometry experiment. M.O. with feedback from T.N., T.V.A.G.O. and S.C.K. carried out the post-experimental data analysis. M.O. performed the simulations using transfer-matrix calculations. G.A.-P., A.Y.N., and P.A.-G. interpreted the results of the theoretical approaches. M.O., T.N., T.V.A.G.O., S.C.K. and L.M.E. prepared the manuscript with input from P.A.-G. All authors took part in the interpretation of the phenomena and contributed to the manuscript.

**Notes**

The authors declare no competing financial interest.

**Acknowledgments**

M.O., T.N., F.H.F., F.G.K., O.H., T.V.A.G.O., S.C.K. and L.M.E. acknowledge the financial support by the Bundesministerium für Bildung und Forschung (BMBF, Federal Ministry of Education and Research, Germany, Project Grant N$^{os}$ 05K16ODA, 05K16ODC, 05K19ODA, and 05K19ODB) and by the Deutsche Forschungsgemeinschaft (DFG, German Research Foundation) under Germany's Excellence Strategy through Würzburg-Dresden Cluster of Excellence on Complexity and Topology in Quantum Matter - ct.qmat (EXC 2147, project-id 390858490). M.O. acknowledges the support form Jiahua Duan in reviewing the manuscript. F.G.K. and L.M.E. gratefully acknowledge the financial support by the DFG through the project CRC1415 (ID: 417590517). G.Á.-P. acknowledges the support through the Severo Ochoa Program from the Government of the Principality of Asturias (grant numbers PA-20-PF-BP19-053). P.A.-G. acknowledges the support from the European Research Council under Consolidator grant no. 101044461, TWISTOPTICS and the Spanish Ministry of Science and Innovation (State Plan for Scientific and Technical Research and Innovation grant number PID2019-111156GB-I00). A.Y.N. acknowledges the Spanish Ministry of Science and Innovation (grant PID2020-115221GB-C42).